\documentstyle[12pt,prl,preprint,aps]{revtex}
\begin{document}
\draft
\title{MISSING AND QUENCHED GAMOW TELLER STRENGTH}

\author{E. CAURIER \thanks{Centre de Recherches Nucl\'eaires,
    IN2P3-CNRS/Universit\'e Louis Pasteur, Division de Physique
    Th\'eorique, BP 20, F-6037 Strasbourg Cedex, France}, A.
  POVES\thanks{Departamento de F\'{\i}sica Te\'orica, Universidad
    Aut\'onoma de Madrid, 28049 Madrid (Spain)} and A. P.
  ZUKER$^*$}

\maketitle

\begin{abstract}
Gamow-Teller strength functions in full $(pf)^{8}$ spaces are
calculated with sufficient accuracy to ensure that all the states in
the resonance region have been populated. Many of the resulting peaks
are weak enough to become unobservable. The quenching factor necessary
to bring into agreement the low lying observed states with shell model
predictions is shown to be due to nuclear correlations. To within
experimental uncertainties it is the same that is found in one
particle transfer  and (e,e') reactions. Perfect consistency between
the observed $^{48}Ca(p,n)^{48}Sc$ peaks and the calculation is
achieved by assuming an observation threshold of 0.75\% of the total
strength, a value that seems typical in several experiments
\end{abstract}

Since the time of the pioneering (p,n) experiments
\cite{good},\cite{gaar},
and the more recent (n,p) ones \cite{vett},\cite{madey} it has been
possible to know the full Gamow Teller strength functions of many
nuclei. The most striking result is that a large fraction of the
theoretically expected sum rules for the $\sigma \tau$ operators,
$S_{+}$ and $S_{-}$, is not visible. The precise amount may be
difficult to asses, in particular because calibration discrepancies
with beta decay measures \cite{adel}, \cite{aufder},  but there is no
doubt that it is substantial and a reduction by a factor 0.6 of
$S_{+}$ and $S_{-}$ is currently accepted as standard. This number is
obtained through two different channels. One is the Ikeda sum rule
$S_{+}$ - $S_{-}$= 3(N-Z), which is model independent provided we do
not introduce non-nucleonic degrees of freedom -and we will not.
Therefore the strength difference cannot be quenched, i.e. suppresed.
It is {\em missing} but it must be somewhere \cite{GRB}.

The other indication comes from the well defined, isolated peaks seen
in $\beta$ decays which are about a factor 0.6 weaker than predicted
by the most accurate shell model calculations available
\cite{BW},\cite{cpzm}. Here we can speak of {\em quenching} because
the data demand it.

In section I we will calculate complete strength functions that
suggest that many states must be unobservable. In section II we
decompose the model independent sum rule in a way that makes apparent
that quenching originates in nuclear correlations. In section III we
give the reasons to expect that only about 50\% of the $S_{-}$ sum
rule for $^{48}Ca$ is observed \cite{and1}.

\vspace{1cm}

{\bf I}. To understand how the strength distributes among daugther
states we rely on the method propossed by Whitehead \cite{white} and
now quite popular \cite{bloom},\cite{haxton},\cite{cpz1}. We work in
the full pf shell with the KB3 interaction \cite{kb},\cite{pz1}, and
obtain an exact eigenstate of the target $|\,i>$ {\em in this model
  space}. Then we define states $|S_{\pm}>=\sigma \tau|\,i>$ whose
norms are the sum rules $S_{\pm}$, and we use them as pivots (i.e.
starting states) in a Lanczos tridiagonal construction. After I
iterations we obtain I+1 eigensolutions and the amplitudes of the
pivot in each of them determine their share of strength. The
situation at I=50 is shown in fig. 1 for  $^{48}Ca(p,n)^{48}Sc$, i.e.
$|\,i>$ is the $(pf)^{8}$ T=4 ground state and the pivot is projected
to keep only T=3 states. The first 4 spikes correspond to converged
eigenstates whose position and strength will not change as I
increases. The others should be viewed as doorways which will split as
we evolve. By construction the first 101 ( i.e. 2I+1) moments of the
exact distribution are given by the spikes. If we were to compare with
experiments with poor resolution but infinite detection power this
would be more than sufficient and we could replace each peak by a
gaussian to obtain a smooth function. Since infinite sensitivity is
not available we have to know in detail how the strength splits and
before comparing with data, eliminate the states below the detection
threshold. In fig. 2 we show how this could be done by pushing to 700
iterations, which guarantees convergence for all the states below 10.5
MeV. We shall examine later plausible values for the threshold.

In $^{48}Sc$ the m-scheme matrices are 1.4$\cdot 10^{5}$ dimensional
and there are 8590 J=1 T=3 states. To guard against numerical errors
each new Lanczos vector must be spin and isospin projected and
orthogonalized with respect to the preceeding ones. On a IBM-3090
about to retire, ANTOINE \cite{antoine} can cope with about 100
iterations per hour for this problem.

To analyze a situation in which the density of levels around the
resonance is much higher, we select the $\beta$ decay of $^{48}Mn$ to
$^{48}Cr$ \cite{seki}, \cite{szeri}, which reaches the region where the
calculated strength with standard quenching is still a factor 2 larger
than the observed one \cite{cpzm}. The J=4 T=1 ground state can go to
J=3,4,5 and T=0,1,2 daughters. Since iterations must be done for each
of these separately and since the m-scheme dimension is now
2$\cdot 10^{6}$, exact calculations with large I can become very
heavy. They were done for  I=45  and it was checked that in the region
 of interest,
configurations $(1f_{7/2})^{8-t} (2p_{3/2},1f_{5/2},2p_{1/2})^{t}$
with $t\leq 3$ are sufficient \cite{cpzm}. The corresponding results
are shown in figs. 3 and 4 for I=50 and 300 which makes clear why we
could be spared the effort of an exact calculation with full
convergence. In the blow-up in fig. 5 we see that 81.5$\%$ of the
strength is distributed among the peaks whose share is less than $1\%$ of
the total. Pushing further the number of iterations and increasing the
size of the spaces could only increase the dilution of much strength into
an unobservable background.

\vspace{1cm}

{\bf II}. To understand the origin of the quenching effect we start by
writting the target eigenstate
$|\,\overline{\i}>$ as a 'dressed' model state $|\,i>$

\begin{equation}
  |\,\overline{\i}>=|\,i>+\sum_{j} |\,j><j\,|\hat{A}|\,i>
\end{equation}

where the j states are outside the model space and $\hat{A}$ is
the correlation operator. Now we separate $\sigma \tau$ as:

\begin{equation}
\sigma \tau = (\sigma \tau)_{m} + (\sigma \tau)_{r}
\end{equation}

where  $(\sigma \tau)_{m}$ contains the contribution of the model
space ( i.e. in our examples pf orbits) and $(\sigma
\tau)_{r}$ contains all others. The total sum rule state can be split
accordingly as:

\begin{equation}
 \frac{\sigma \tau_{\alpha}}{<\overline{\i}\,|\,\overline{\i}>}
 |\,\overline{\i}>  =
   |s_{\alpha}> = |s_{\alpha m}> + |s_{\alpha r}> , \, \, \, \alpha = \pm
\end{equation}

By using exactly the same arguments that lead to Ikeda's sum rule
we have:

\begin{equation}
  S_{-} - S_{+} = 3(n_{m}-z_{m}) +  3(n_{r}-z_{r}) =
(S_{-}  - S_{+})_{m} + (S_{-} - S_{+})_{r},
\end{equation}

where $n_{m}$, $z_{m}$, $n_{r}$ and $z_{r}$ are expectation values of
number operators, for which obviously $n_{m} + n_{r} = N$ and
 $z_{m} + z_{r} = Z$.

  Intuitively it is clear  $(\sigma \tau)_{m}|\,\overline{\i}>$ is a
  state in the model space for the daughter nucleus while
  $(\sigma \tau)_{r}|\,\overline{\i}>$ will produce one outside
  that space. The result is true in leading order of perturbation
  theory and we propose it as a good approximation. ( To be more
  precise demands information about $\hat{A}$  \cite{pz2}).

The consequences are very pleasing because now eq.(4) can be
interpreted as a clean separation of two contributions: one from the
model space and one from outside. The first is then:

\begin{equation}
 (S_{-}-S_{+})_{m} =
 3<\overline{\i}\,|\hat{n}_{m}-\hat{z}_{m}|\,\overline{\i}> =
 3<i\,|\hat{n}_{m}-\hat{z}_{m}|\,i> \cdot \, (0.7)
\end{equation}

(we use $\hat{n}$ and $\hat{z}$ to distinguish operators
from expectation values)

The factor 0.7 comes from the (d,p) data of Vold {\em et al}
\cite{vold} and is consistent with the occupancies near the Fermi
level obtained in (e,e') scattering \cite{cav},\cite{pan},\cite{ben}.
Quenching therefore originates in deep correlations that reduce by
about 40$\%$ the discontinuity at the Fermi surface.

Although the precise form of the renormalized $\sigma \tau$ operator
acting in the model space is in principle complicated, to satisfy
eq.(4) it is sufficient to use $\sqrt{0.7} (\sigma \tau)_{m}$, which
is standard practice, except that the factor is in general
$\sqrt{0.6}$.  In view of experimental and theoretical uncertainties
the two factors are most probably compatible. Furthermore, these
arguments establish nuclear correlations as entirely responsible for
quenching, again within uncertainties.

The inequality $ S_{-} = 3(N-Z)+ S_{+} \geq 3 (N-Z) $ is often used to
establish the discrepancy between the measured $S_{-}$ and the
theoretical bound, and it is also argued that $S_{+}$ is likely to be
small in nuclei of large neutron excess, when it originates in
correlation terms (i.e. r-components outside the model space).
In such cases $S_{+}$ may be difficult to measure rather than small
and the same could be said of the term $(S_{-}-S_{+})_{r}$ in eq.(4),
always enterely due to correlations. Therefore we propose a statement
that is both consistent with the sum rule and with observations

\begin{center}
              $(S_{-}-S_{+})_{m}= (0.7) \cdot \, 3(N-Z) $
\end{center}

where m stands conveniently for {\em model} and {\em measured}.

It is seen that the quenching and missing strength factors are
identical if we assume that all strength due to correlations is
missing and all strength coming from the model space is measured. As
we shall see    now, experiments probably miss most of the former and
substantial amounts of the latter.

\vspace{1cm}

{\bf III}. To decide which is the observation threshold for
$^{48}Ca(p,n)^{48}Sc$  we note that in the data of Anderson {\em et
  al} \cite{and1} the strong isolated peak at 2.52 MeV (2.3 in figs.
 1,2) collects a strength of 6.8 against 38.7 for all the states in the
interval 4.5-14.5 MeV. The ratio of the two numbers is 5.7 while we
would find 8.4 from fig. 2. In table 1 we show the amount of surviving
strength as a function of the threshold. Selecting a cutoff of
0.75$\%$ of the total $S_{-}$ we obtain fig.  6 where the ratio is now
5.7. It is interesting to compare with fig. 1, which is the one we
would have normally kept, and whose relatively modest lowest peak
becomes now the largest, in line with what is seen experimentally
\cite{and1}, \cite{zhao}. The reader is invited to check (or to
believe) that the two
smallest isolated, observed bumps correspond exactly with the peaks at
3.5 and 5.5 MeV that have (barely) survived the cutoff. This is a very
direct indication that the threshold chosen is indeed realistic. In
addition to these bumps, the data show three gross structures centered
at 7.5, 10 and 12 MeV that correspond closely to what is seen in
fig. 6 (ref \cite{zhao} contains a good plot of the original data
\cite{and1}).

{}From the arguments we have presented, it follows that in the
$^{48}Ca(p,n)^{48}Sc$ experiment, some 25$\%$ of the model strength
goes unobserved. It is quite plausible to assume that the strength
associated to correlations will be spread among smallish peaks and
that few of them will survive the cutoff. A $^{48}Ca(n,p)^{48}K$
experiment will be very welcome even if very little is seen, to
confirm the $S_{-}-S_{+}$ loss factor (0.6-0.7)(0.75) i. e. some
$50\%$.  It is important to note that this value is not expected to be
typical. In fig. 5 we find that only 19$\%$ of the strength is located
in peaks that survive a 1$\%$ cutoff and even the more generous
0.5$\%$ will only spare 35$\%$ of the strength. Such situations will
arise whenever the resonance moves to regions of high level density
where dilution is severe. Conversely, for low level densities we may
recover the standard factor. It is also of interest to stress that
there is no reason to expect that a pile up of small peaks below
threshold in a narrow bin of energy could produce a legitimate signal.
In this sense, tails and background should be excluded from the
collection of strength unless good reasons could be given that they
contain large enough peaks.  A remark worth doing in this context is
that the smallest isolated peak detected in any measure we have
consulted contains approximately 1$\%$ of the total strength
\cite{miscsd}.

\vspace{0.5cm}

To conclude. The quenching problem can be solved by invoking deep
correlations. The result was either known, or suspected or believed
\cite{bertch} but we think the simple proof presented here may be new
and has the advantage of relating the standard factor $\sqrt{0.6}$ to
occupancies at the Fermi level. (Again, this was probably guessed by
somebody). Missing strength is another matter. It is certainly there,
mostly in the region of the resonance (for model states at least) but
its identification will demand an extra effort. Since experimentalist
have done already quite extraordinary things it may be unfair -but not
hopeless- to demand more.

\vspace{2cm}

\acknowledgements

This work has been partly supported by the
IN2P3(France)- CICYT(Spain) agreements and by
DGICYT(Spain) grant PB89-164.

\newpage

\begin{table}[1]
\begin{center}
\leavevmode
\begin{tabular}{|cccc|}
\hline
cutoff($\%$) & I=50 & I=300 & I=700 \\
\hline
0.05 & 99.99 & 99.14 & 97.85 \\
0.10 & 99.94& 98.57 & 96.43 \\
0.20 & 99.45 &97.10 & 93.15  \\
0.50 & 97.92 & 89.76 & 83.55 \\
0.75 & 97.35 & 86.98 & 73.37 \\
1.00 & 96.36 & 81.76 & 67.95 \\
\hline
\end{tabular}
\end{center}
\caption{$^{48}Ca (p,n) ^{48}Sc$: percentage of strength located in
  peaks whose share of the total strength is larger than the cutoff,
  as a function of the number of Lanczos iterations.}
\end{table}



\begin{figure}
  \caption{$^{48}Ca  (p,n) ^{48}Sc$ Gamow Teller strength function:
    50 iterations Lanczos. B(GT) in \% of the total strength.}
  \label{fig:fig1}
\end{figure}

\begin{figure}
  \caption{$^{48}Ca  (p,n) ^{48}Sc$ Gamow Teller strength function:
  700 iterations Lanczos. B(GT) in \% of the total strength.}

  \label{fig:fig2}
\end{figure}

\begin{figure}
  \caption{$^{48}$Mn $\rightarrow$  $^{48}$Cr Gamow Teller strength
    function: 50 iterations Lanczos. B(GT) in \% of the total
    strength.}
  \label{fig:fig3}
\end{figure}

\begin{figure}
  \caption{$^{48}Mn \rightarrow  ^{48}Cr$ Gamow Teller strength
    function: 300 iterations Lanczos. B(GT) in \% of the total
    strength.}

  \label{fig:fig4}
\end{figure}

\begin{figure}
  \caption{$^{48}Mn \rightarrow  ^{48}Cr$ Gamow Teller strength
    function: 300 iterations Lanczos. B(GT) in \% of the total
    strength.
    Only states carrying less than 1 \% of the total
    strength are plotted.}
  \label{fig:fig5}
\end{figure}

\begin{figure}
  \caption{$^{48}Ca  (p,n) ^{48}Sc$ Gamow Teller strength function:
    700 iterations Lanczos. B(GT) in \% of the total strength. Only
    states carrying more than 0.75 \% of the total strength are
    plotted.}
  \label{fig:fig6}
\end{figure}


\begin{thebibliography}{99}

\bibitem{good} C.D. Goodman, Nucl. Phys. A374, 241c (1982): C.D.
Goodman {\em et al }, Phys. Rev Lett. 44, 1755 (1980).

\bibitem{gaar} C. Gaarde {\em et al }, Nucl. Phys. A334, 248 (1980).

\bibitem{vett} M.C. Vetterly {\em et al }, Phys. Rev. Lett. 59, 439
(1987): Phys. Rev. C40, 559 (1989).

\bibitem{madey} R. Madey   {\em et al }, Phys. Rev. C35, 2011
(1987): C36, 1647 (1987).

\bibitem{adel} E.G. Adelberger  {\em et al }, Phys. Rev. Lett. 67,
3658 (1991).

\bibitem{aufder} M. B. Aufderheide  {\em et al }, Phys. Rev. C46, 2251 (1992).

\bibitem{GRB} C.D. Goodman, J. Rapaport and S.D. Bloom, Phys. Rev.
C42, 1150 (1990).

\bibitem{BW} B.A. Brown and B. H. Wildenthal, At. Data Nucl. Data.
 Tables 33, 347 (1985).

\bibitem{cpzm} E. Caurier, A. Poves , A. Zuker and G. Martinez-Pinedo,
               preprint FTUAM-93/01, submitted to Phys. Rev. C.


\bibitem{and1} B.D. Anderson  {\em et al }, Phys. Rev. C31, 1161 (1985).

\bibitem{white} R.R. Whitehead, in 'Moment methods in many fermion
systems', eds. B.J. Dalton, S.M. Grimes, J.D. Vary and S.A. Williams
(Plenum, New York, 1980) p.235.

\bibitem{bloom} S.D. Bloom and G.M. Fuller, Nucl. Phys. A440, 511 (1985).

\bibitem{haxton} J. Engel, W.C. Haxton and P. Vogel, Phys. Rev. C46,
2153 (1992).

\bibitem{cpz1} E. Caurier, A. Poves and A. Zuker, Phys. Lett. 252B, 13 (1990).

\bibitem{kb} T.T.S. Kuo and G.E. Brown, Nucl. Phys. A114,241 (1968).

\bibitem{pz1} A. Poves and A. Zuker, Phys. Reports 70, 235 (1981).

\bibitem{antoine} E. Caurier, code ANTOINE, Strasbourg 1989.

\bibitem{seki} T. Sekine  {\em et al }, Nucl. Phys. A467, 93 (1987).

\bibitem{szeri} J. Szeripo  {\em et al }, Nucl. Phys. A528, 203 (1991).

\bibitem{pz2} A. Poves and A. Zuker, Phys. Reports 71, 141 (1991).

\bibitem{vold} P.B. Vold  {\em et al }, Nucl. Phys. A302, 12 (1978).

\bibitem{cav} J.M. Cavedon  {\em et al }, Phys. Rev. Lett. 49, 978 (1982).

\bibitem{pan} V. R. Pandharipande, C.N. Papanicolas and J Wambach,
Phys. Rev. Lett. 53, 1133 (1984).

\bibitem{ben} O. Benhar, V. R. Pandharipande and S.C. Pieper, Rev.
Mod. Phys. 65, 817 (1993).

\bibitem{zhao} L. Zhao, B. A. Brown and W.A. Richter, Phys. Rev. C42,
1120 (1990)

\bibitem{miscsd} B.D. Anderson   {\em et al }, Phys. Rev. C36, 2195
(1987): C43, 50 (1991).

\bibitem{bertch} G.F. Bertsch and H. Esbensen, Rep. Prog. Phys. 50,
607 (1987).

\end{thebibliography}
\end{document}